 \documentclass[10pt]{article}

\usepackage{mypaper}
\usepackage{verbatim}
\usepackage{algorithm}
\usepackage{algorithmic}

\usepackage{listings}
\usepackage{color}
\newif\ifexploitdetail
\exploitdetailtrue

\title{Incomplete Sets in $\P$ under log-space Reduction}
\ifdefined\lipics
\author{Reiner Czerwinski}{TU Berlin (alumnus)}{}{}{}
\authorrunning{Reiner Czerwinski}
\Copyright{Reiner Czerwinski}
\keywords{ L versus P, Complexity Classes, P-completeness, inherently sequential}
\ccsdesc{Theory of computation~Complexity classes}
\else
\ifsnjnl
\author*{\fnm{Reiner} \sur{Czerwinski}\email{Reiner.Czerwinski@posteo.de}}
\affil{No Institution}
\else
\author{Reiner Czerwinski}
\fi
\fi

\newcommand{\bsp}[1]{\ensuremath{B_{#1}}}
\newcommand{\usp}[1]{\ensuremath{U_{#1}}}
\newcommand{\Ap}{\ensuremath{\mathcal{A}}}

\begin{document}
\begin{headdata}
  \begin{abstract}
   \ In this article, we construct $\P$-incomplete sets under log-space many-one reduction.
    For a positive integer $k$, $\DSPACE(\log^{k+1}(n))$-complete sets
    are not log-space reducible to sets contained in $\DSPACE(\log^k(n))$
    by the space hierarchy theorem.
    
    It is generally undecidable whether a set is in $\P$.
    Hence, according to the space hierarchy theorem, 
    there exists a set in $\DTISP(\poly,\log^{k+1}(n))$,
    which is not log-space reducible to sets contained in
    $\DTISP(\poly,\log^{k}(n))$.

    Thus, no set in $\SC$ is $\P$-complete,
    which implies that $\P$ is not equal to $\cL$, and to $\NC$.
    Therefore, complete sets in $\P$ are inherently sequential.
  \end{abstract}
\end{headdata}

\section{Introduction}
The set \[\Ap=\{(M,x,1^t) \mid \text{ TM }M \text{ accepts } x \text{ within } t \text{ steps} \} \] is $\P$-complete
under log-space many-one reduction~\cite[page 176, Th. 7.24]{Homer2011}, i.e.\
$\Ap\in\P$, and for every set $X$ in $\P$, $X\le_m^{\log}\Ap$.

It is generally undecidable whether a TM accepts an input.
However, when a TM accepts an input, it does this so finitely many steps.
Therefore,
\begin{equation}\label{eq:acceptfinite}
 x \in L(M) \iff \exists t\in\nat\;(M,x,1^t) \in \Ap
\end{equation}
for all $M$ and $x$. Hence,
we can apply computability theory to analyze the $\P$-complete set $\Ap$.

In this paper, we construct $\P$-incomplete sets.
The sets
\begin{equation}\label{eq:usp}
  \begin{aligned}
       \usp{k} := \{(M,x,1^s) \mid &\text{ The TM } M 
    \text{ accepts }x
    \text{ within }\log^k(s) \text{ tape space,}\\
   &\text{ and the runtime of }M\text{ is polynomially bounded}\\&\text{ in }|x|+s.
   \}
  \end{aligned}
\end{equation}
are in $\P$, since for each positive integer $k$, $\usp{k}\in\DTISP(\poly,\log^k(n))$.

Unfortunately, these sets are hard to analyze directly, as shown
in Section~\ref{spacevstime}. 
For this reason, we observe the sets:
\begin{equation}\label{eq:bsp}
  \bsp{k} :=  \{(M,x,1^s) \mid \text{ TM } M \text{ accepts } x
             \text{ within } \log^k(s) \text{ tape space} \}.
\end{equation}
           
For each positive integer $k$, the set $\bsp{k}$ is 
$\DSPACE(\log^k(n))$-complete under log-space many-one reduction.

\begin{lemma}\label{red}
$\usp{k} \le_m^{\log} \bsp{k}$ and $\usp{k} \le_m^{\log} \Ap$
\end{lemma}
\begin{proof}
    If $(M,x,1^s) \in \usp{k}$ then  $(M,x,1^s) \in \bsp{k}$.
    Therefore, $\usp{k} \le_m^{\log} \bsp{k}$.

   For all $k\in\nat$,
   $\usp{k}\in\P$ and $\Ap$ is $\P$-complete.
   Thus, $\usp{k} \le_m^{\log} \Ap$.
\end{proof}




In Section~\ref{sectionincomplete}, we prove
that $\bsp{k+1}\not\le_m^{\log} \bsp{k}$ for $k\in\nat$ in Theorem~\ref{reduc}.
Theorem~\ref{reduc} and Lemma~\ref{undecidable} imply
that $\usp{k+1}\not\le_m^{\log} \usp{k}$ for $k>0$.
Therefore, all $\usp{k}$ are $\P$-incomplete.

To prove irreducibility, we need methods from computability theory.
To emphasize this, we put Lemma~\ref{undecidable} in its own section,
Section~\ref{spacevstime}.
We prove with the halting problem in Lemma~\ref{undecidable}
that it is undecidable whether a set is in $\P$,
and therefore, whether a set in $\DSPACE(\log^k(n))$ is in
$\DTISP(\poly,\log^k(n))$ for~$k>1$.

In Section~\ref{conclusion}, we show other $\P$-incomplete problems, and that
there is no parallel algorithm with efficient speedup for any
$\P$-complete problem.

Baker, Gill, and Solovay have proved the  relativization barrier for P versus NP~\cite{baker1975relativizations}.
Any proof separating L from P also must not relativize.
In Section~\ref{relativ}, we explain why the proof in our paper circumvents the relativization barrier.

\section{On Computability}\label{spacevstime}
\ifexploitdetail
\newcommand{\testTM}{\ensuremath{M'}}
\else
\newcommand{\testTM}{\ensuremath{M}}
\fi

%

As mentioned in Eq.~(\ref{eq:acceptfinite}),
we use computability to analyse the complexity class~$\P$.
\begin{lemma}\label{undecidable}
  Let $k>1$.
  For a TM with space requirement of $\Theta(\log^k)$,
  no algorithm can decide in general
  whether the runtime of the TM is polynomial.
\end{lemma}
\begin{proof}
  We reduce the halting problem~\cite{turing1936computable} to the question
  of whether the runtime is polynomial.

  Given a TM $M$, which contains a read-only input tape
  and a limited work tape with read-write head,
  and an input $x$.
  The
  \ifexploitdetail{TMs $M,$ and $\testTM$ terminate}
  \else{TM $M$ terminates}
  \fi
  if
  \ifexploitdetail
  they reach
  \else
  it reaches
  \fi
  an accepting state, or
  rejecting state.
  For convenience, we assume that \ifexploitdetail{ each of the TMs }
  \else{ the TM $M$ }\fi only reaches the
  rejecting state if it requires more space than reserved
  on the work tape during the runtime.
  
  We construct a multi-tape TM
  as described in Algorithm~\ref{slowTM}.
  \newcommand{\multiTM}{\ensuremath{\text{testLM\_maybe\_poly}}}
  \begin{algorithm}
    \ifexploitdetail
    \caption{FUNCTION $\multiTM(k,M,\testTM,x)$ \{Polynomial runtime iff TM $\testTM$ accepts itself as input\}}
    \else
    \caption{Polynomial runtime iff TM $\testTM$ accepts itself as input}
    \fi
    \begin{algorithmic}\label{slowTM}
      \REQUIRE{TM $M$ with input $x$, and $k>1$ }
      \REQUIRE{TM $\testTM$ with input $\langle \testTM\rangle$ }
      
      \ENSURE{The string $x$ is on the first, and $\testTM$ on the second read-only tape}
      \ENSURE{three tapes with read-write head; two work tapes and a counter tape}

      \STATE{allocate $\log^k(|x|)$ space on each of the tapes with read-write head with '0'-symbols}
      \STATE{ move read-write heads to the middle of the allocated space}
      \ENSURE{TM $M$ rejects if the read-write head of its work tape leaves the allocated space}
      \WHILE{counter contains a '0'-symbol}
      \STATE{add 1 in binary to the counter}
      \IF{$M$ has not terminated on input $x$}
      \STATE{calculate one step of TM $M$ on input $x$}\COMMENT{with first work tape }
       \ENDIF
      \IF{$\testTM$ has not terminated on input $\langle \testTM\rangle$}
      \STATE{calculate one step of TM $\testTM$ on input $\langle \testTM\rangle$}\COMMENT{with second work tape }
      \IF{$\testTM$ has accepted input $\langle \testTM\rangle$}
      \STATE{ resize the counter tape to $\log(|x|)$ cells}
      \STATE{fill counter with '0'-symbols }
      \ENDIF
      \ENDIF

       \ENDWHILE
     \RETURN{the state of $M$ on input $x$}
     \COMMENT{accept or not}
   \end{algorithmic}
          \end{algorithm}

  This multi-tape TM runs simultaneously $M$ on input $x$
  and $\testTM$ on input $\langle \testTM\rangle$ using different work tapes.
  The space of each work tape is limited to $\log^k(|x|)$.
  Additionally, there is a counter tape initialized with
  a string of $\log^k(|x|)$ '0'-symbols.

  While the counter contains a '0'-symbol,
  we do the following steps:

  We increment the counter in binary.
  Then we execute one step of $M$ on input $x$ and one step of $\testTM$ on the encoding of
  itself as input.
  As soon as the computation of $\testTM$ on input $\langle \testTM\rangle$ reaches an accepting state,
  we resize the counter to $\log(|x|)$ cells filled with '0'-symbols.

           If the TM $\testTM$ accepts the encoding of itself, then Algorithm~\ref{slowTM} executes
          $O(2^{\log(|x|)})=O(|x|)$ iterations. Otherwise, the number of
          iterations is $2^{\log^k(|x|)}= |x|^{\log^{k-1}(|x|)}$.
    The multi-tape TM has a polynomial runtime if and only if
    the TM $\testTM$ accepts the encoding of itself, i.e.\ $\langle \testTM \rangle \in L(\testTM)$.
    Therefore, this property is undecidable in general.
\end{proof}

The following corollary follows directly from Lemma~\ref{undecidable}:
\begin{corollary}\label{polyundecide}
  For $k>1$, it is undecidable for a triple in $B_k$
  whether this triple is contained in $U_k$.
\end{corollary}

\section{Space Hierarchy within \P}\label{sectionincomplete}
Recall the sets $\bsp{k}$ in Eq.~(\ref{eq:bsp}),
and $\usp{k}$ in Eq.~(\ref{eq:usp}).


\begin{theorem}\label{reduc}
  For each positive integer $k$, the set $\bsp{k}$ is
  $\DSPACE(\log^k(n))$-complete under log-space many-one reduction.
\end{theorem}
\begin{proof}
  Obviously, $\bsp{k}\in\DSPACE(\log^k(n))$.

  If $X \in\DSPACE(\log^k(n))$, then $X$ is the language of a TM
  that requires $O(\log^k(n))$ tape space.
  This means there is a constant $C>0$, such that the tape space required 
  by the TM for an input $x$ is less than $C\cdot\log^k(|x|)$.
   Using a log-space transducer,
  we can construct a triple consisting of the TM, the input $x$, and a padding
    argument of length $C\cdot\log^k(|x|)$.
    This triple is in $\bsp{k}$ if and only if the TM accepts the
    input $x$.
\end{proof}
\begin{theorem}\label{noreduce}
  If $k$ is a positive integer,
  then
  $\bsp{k+1} \not\le_m^{\log} \bsp{k}$.
\end{theorem}
\begin{proof}
  Due to the space hierarchy theorem~\cite{HLS65}, there exists a set $X \in \DSPACE(\log^{k+1})$ such that
  $X \not\in \DSPACE(\log^k)$.
  Thus, by Theorem~\ref{reduc}, $X \le_m^{\log} \bsp{k+1}$ and $X \not\le_m^{\log} \bsp{k}$.
  This implies $\bsp{k+1} \not\le_m^{\log} \bsp{k}$.
\end{proof}

\begin{lemma}
  If $k$ is a positive integer,
  then
  $\usp{k+1} \not\le_m^{\log} \bsp{k}$.
\end{lemma}
\begin{proof}
  To construct a contradiction, we assume that $\usp{k+1} \le_m^{\log} \bsp{k}$.
  In this case, there would be a general log-space transducer mapping
  the triple $(M,x,1^s)\in\usp{k+1}$ to an element in $\bsp{k}$.
  By Theorem~\ref{noreduce},
  there is no such transducer for a triple $(M,x,1^s)\in\bsp{k+1}$
  to $\bsp{k}$.
  The construction of this transducer could be used as an exploit to
  decide whether the triple in $\bsp{k+1}$ is also in $\usp{k+1}$.
  However, by Corollary~\ref{polyundecide}, whether this triple is contained
  in $\usp{k+1}$ is generally undecidable.

  \ifexploitdetail
  We assume that there is a log-space transducer $T$ from
  $\usp{k+1}$ to $\bsp{k}$.
  By Theorem~\ref{noreduce}, there is a TM $M$ with space limit $\log^{k+1}(n)$, such that the transducer $T$ fails to map $M$ to a TM with space limit $O(\log^{k}(n))$.
  Without loss of generality there is an input $x\in L(M)$, but $x\not\in L(T(M))$.

  We construct an algorithm, denoted Algorithm~\ref{enuminput}, with TM $M$,
  which is irreducible via $T$, and an arbitrary but fixed TM $\testTM$.

  Algorithm~\ref{enuminput}
  tests for each input
  whether Algorithm~\ref{slowTM} returns the same result with and
  without applying transducer $T$, and halt, if both results are unequal.
  
  If the TM $\testTM$ does not accept itself as input, then Algorithm~\ref{slowTM}
  returns true, if and only if $M$ accepts $x$ with space limit $\log^{k+1}(|x|)$.
  Thus, there exists an input $x$, where  Algorithm~\ref{slowTM} accepts
  $x$, but not after applying the transducer $T$.

  If the TM $\testTM$ accepts itself as input, then Algorithm~\ref{slowTM}
  has polynomial run time. Hence, the transducer $T$ works correctly.
  Applying $T$ on Algorithm~\ref{slowTM} does not change the result.

  \newcommand{\alf}{\ensuremath{\Sigma}}
  \newcommand{\multiTM}{\ensuremath{\text{testLM\_maybe\_poly}}}
  \newcommand{\params}{\ensuremath{(k,M,\testTM,x)}}
  \newcommand{\multiTMpar}{\ensuremath{\multiTM(k,M,\testTM,x)}}

  \begin{algorithm}
   \caption{find input that Algorithm~\ref{slowTM} classifies different after transducing}
    \begin{algorithmic}\label{enuminput}
      \REQUIRE{TMs $M,\testTM$, and $k>1$ }
      \ENSURE{ input alphabet of $M$ is $\alf$ }
      \ENSURE{ TMs $M,\testTM$ fulfill requirements in Algorithm~\ref{slowTM} }
      \STATE{}
      \COMMENT{function $\multiTM$ is described in Algorithm~\ref{slowTM} }
      \FORALL{\begin{math}x\in\alf^*\end{math}}
      \IF{\begin{math}\multiTMpar \not= T(\multiTMpar)\end{math}}
      \STATE{halt}
      \ENDIF
      \ENDFOR
    \end{algorithmic}
  \end{algorithm}
  
  Thus, Algorithm~\ref{enuminput}
  terminates if and only if
  $\testTM$ does not terminate with itself as input.
  This would allow us to decide the halting problem, and so,
  the transducer $T$ cannot exist.
  
  \fi
  
\end{proof}
\begin{theorem}\label{polylog}
  If $k$ is a positive integer,
 then
   $\usp{k+1} \not\le_m^{\log} \usp{k}$.
\end{theorem}
\begin{proof}
   We have
   $\usp{k+1} \not\le_m^{\log} \bsp{k}$, but $\usp{k} \le_m^{\log} \bsp{k}$.
\end{proof}

\section{Conclusion}\label{conclusion}
For all $k$, $\usp{k}\in\P$ and $\Ap\not\le_m^{\log} \usp{k}$, so $\usp{k}$ is $\P$-incomplete.
From the existence of incomplete sets, it follows:
\begin{corollary}
$\cL \neq \P$
\end{corollary}
\begin{proof}
  If $X\in\cL$, then $X\le_m^{\log}\usp{2}$. Therefore, $\Ap\not\le_m^{\log}X$,
  since $\Ap\not\le_m^{\log}\usp{2}$. So, $\Ap\in\P$, and $\Ap\not\in\cL$.
\end{proof}
The class $\SC$ is defined as
\[
 \SC = \bigcup_{k\in\nat} \DTISP\bigl(\poly,\log^k(n)\bigr).
\]
\begin{corollary}
  $\SC \neq \P$
\end{corollary}
\begin{proof}
  For every $X\in\SC$, there exists $Y\in\SC$ such that $X\not\le_m^{\log}Y$.
  Thus, there is no $\P$-complete set in $\SC$.
\end{proof}

\begin{corollary}\label{nc}
  $\NC \neq \P$
\end{corollary}
\begin{proof}
  By definition, $\NC = \bigcup_{i\in\nat}\NC^i$, and
  $\NC^i \subseteq \DSPACE(\log^i(n))$~\cite{Borodin1977OnRT}, so
  there is no $\P$-complete set in $\NC$.
\end{proof}
Corollary~\ref{nc} implies
that $\P$-complete problems are inherently sequential~\cite{greenlaw1995limits}.
\section{Notes on the Relativization Barrier}\label{relativ}
To separate $\cL$ and $\P$, the relativization barrier applies as it did for $\P$ versus $\NP$~\cite{baker1975relativizations}.

Aaronson and Wigderson wrote~\cite{algebrization}:
``On the other hand, if we allow only polynomially-long queries,
then results based on padding -- for
example, $\P = \NP \Longrightarrow \EXP = \NEXP$ -- will generally fail to relativize.''

The proof method in this paper uses padding, and the oracle queries are bounded
by the input length including the padding argument. Thus, the method is suitable to circumvent the relativization
barrier\footnote{This phrase is self-plagiarism because it is used in many papers of the author since 2020}.
\section*{Acknowledgments}
Many thanks to Lance Fortnow. His email correspondence helped me find bugs in the predecessor to this article~\cite{czerwinski2021separation}.
The bugs were also mentioned by Ian Clingerman, and Quan Luu from the University of Rochester~\cite{clingerman2023critiqueczerwinskisseparationrm}.   
\bibliography{lit}{}
\ifsnjnl
\bibliographystyle{sn-aps.bst}
\else
\bibliographystyle{plain}
\fi

 \end{document}